\documentclass[preprint,nofootinbib,a4paper,11pt]{revtex4}

\usepackage[centertags]{amsmath}
\usepackage{amsfonts}
\usepackage{amssymb}
\usepackage[sort&compress]{natbib}% упорядочивает ссылки
\usepackage[hypertex]{hyperref}%hypertex
\usepackage[dvips]{graphicx}% рисунки

\DeclareMathOperator{\Sp}{Sp}
\DeclareMathOperator{\gh}{gh}
\DeclareMathOperator{\sign}{sign}
\DeclareMathOperator{\re}{Re}
\DeclareMathOperator{\im}{Im}
\DeclareMathOperator{\arccot}{arccot}
\DeclareMathOperator{\arccoth}{arccoth}

\newcommand{\spx}{\mathbf{x}}
\newcommand{\spy}{\mathbf{y}}

\newcommand{\e}{\varepsilon}
\newcommand{\vf}{\varphi}
\newcommand{\s}{\sigma}

\newcommand{\be}{\beta}
\newcommand{\ga}{\gamma}
\newcommand{\Ga}{\Gamma}
\newcommand{\de}{\delta}
\newcommand{\De}{\Delta}

\newcommand{\la}{\lambda}
\newcommand{\La}{\Lambda}
\newcommand{\ups}{\upsilon}

\begin{document}

\setlength{\unitlength}{1mm}% устанавливает единицу длины в окружении picture и пр.

\title{One-loop effective potential of the Higgs field on the Schwarzschild background}

\date{\today}

\author{P.O. Kazinski}

\email{kpo@phys.tsu.ru}

\affiliation{Physics Faculty, Tomsk State University, Tomsk, 634050 Russia\\
Institute of Monitoring of Climatic and Ecological Systems, SB RAS, Tomsk, 634055 Russia}

\begin{abstract}

A one-loop effective potential of the Higgs field on the Schwarzschild background is derived in the framework of a toy model: a SO(N) scalar multiplet interacting with the gauge fields, the SO(N) gauge symmetry being broken by the Higgs mechanism. As expected, the potential depends on the space point and results in a mass shift of all massive particles near a black hole. It is shown that the obtained potential depends on the space point through the metric component $g_{00}$ in the adapted coordinates and has the same form for an arbitrary static, spherically symmetric background. Some properties of this potential are investigated. In particular, if the conformal symmetry holds valid for massless particles on the given background, there exist only two possible scenarios depending on a sign of an arbitrary constant arising from the regularization procedure: the masses of all massive particles grow infinitely when they approach the black hole horizon, or the gauge symmetry is restored at a finite distance from the horizon and all particles become massless. If the conformal symmetry is spoiled, an additional term in the effective potential appears and the intermediate regime arises. Several normalization conditions fixing the undefined constants are proposed, and estimations for the mass shifts are given in these cases.

\end{abstract}

\maketitle

%\newpage
\section{Introduction}

There is a belief that quantum effects are only relevant for the black hole physics when the Compton wave-length $\hbar/Mc$ of the black hole is of the order of its gravitational radius $2GM/c^2$. This opinion is mainly based on the study of the Hawking radiation \cite{Hawk}, which is incredibly small for regular black holes, or just on dimensional reasons. At the same time, we know that the metric field has a singularity on the horizon of a black hole, at least, in the Schwarzschild coordinates. So, we can expect a new physics in its neighborhood, which is not catched by dimensional considerations since there are large dimensionless parameters -- the metric components. The aim of this paper is to derive the effective potential of the Higgs field on the Schwarzschild background, which governs the masses of particles in the standard model. It is evident that the effective potential must depend on a point of the space. Moreover, it is almost evident that it should be singular at the horizon of a black hole as the metric components are singular. So, we expect that the masses of particles considerably change near a black hole.

In the seminal paper by Coleman and Weinberg \cite{ColWein}, it was suggested that quantum fluctuations may change the effective action in such a way that a spontaneous breaking of the gauge symmetry occurs. That is, quantum fluctuations change a landscape of the potential entering the effective Lagrangian, and the properly constructed perturbation theory for a classically massless gauge invariant model will contain massive vector bosons. Investigating below the effective potential of the Higgs field on the Schwarzschild background, and, more generally, on an arbitrary static, spherically symmetric space-time, we shall see that the contributions to it due to quantum fluctuation are of the same form as in the Coleman-Weinberg potential, but with coefficients depending on a point of space. It turns out that, in interpolating the one-loop results, quantum fluctuations are so strong near the horizon of a black hole that they can restore spontaneously broken symmetry of the standard model or make massive particles to be very massive with infinite masses on the horizon. More rigorously, if a conformal symmetry holds valid for massless particles on the considered background then there only exist two scenarios depending on a sign of the undetermined constant coming from a regularization procedure. In the first case, the symmetry is restored and all particles become massless at a finite distance from the horizon of a black hole, while in the second case the masses of all massive particles grow, when one approaches the horizon, and tend to infinity. These two scenarios can be called as the ``hot'' and ``cold'', respectively, as the matter falling into it becomes hot in the first case and cold in the second one. If the conformal symmetry is spoiled, the additional constant governing the mass shifts arises. Properly chosen, it can compel to tend the vacuum expectation value of the Higgs field to a finite nonzero value on the horizon. So, the intermediate regime arises. It is clear that the mass shift results in many unusual physical effects and we do not enlarge on them. The effective potential also allows us to find how the energy density of the Higgs field varies from the point to point in a static, spherically symmetric space-time. However, to this end we have to fix additionally one arbitrary constant.

Deriving the effective potential we, of course, shall make certain approximations. First, we use a toy model instead of the standard one. We consider a scalar $SO(N)$ multiplet interacting with the gauge fields taking values in the Lie algebra $so(N)$. The gauge symmetry is spontaneously broken by the Higgs mechanism down to $SO(N-1)$. In the upshot, we choose $N=4$ in order to have three massive vector bosons as in the electroweak sector of the standard model. Second, we shall find the one-loop correction to the effective action only. In other words, we assume that the model is in a perturbative regime, i.e., the energies are sufficiently high and the coupling constants are small. This enables us to cast out the higher order terms of the loop expansion. Third, calculating the determinant emerging from an integration of the Gaussian path-integral, we exploit a quasiclassical (short-wave) approximation. That is, we take effectively into account the modes with the wave-lengths much smaller than the gravitational radius $r_g$. The contribution of the long-wave modes to the effective action is well approximated by the renormalized classical action and we also add it to the final result. In fact, we add the kinetic term only neglecting a self-interaction of the Higgs field by means of the gauge fields. Notice that the quasiclassical approximation allows us not to make any distinction between the one-loop contributions to the effective action of one scalar particle and one polarization of a vector boson regardless of the fact that they interact differently with the gravitational field. Calculating the functional determinant, we do not use the standard method of the heat kernel expansion \cite{Schwing,DeWittDTGF,VasilHeatKer}. It gives an asymptotic series in the powers of $1/m$, where $m$ is a mass of the particle, which is proportional to the Higgs field for the vector bosons. The higher coefficients of this series are unknown, while to take a few first terms from it is unacceptable for our goal to obtain a regular effective potential of the Higgs field.

%Besides, these quasiclassical calculations are easily generalized from the Schwarzschild background to any static, spherically symmetric metric.

The paper is organized as follows. In Sec. \ref{eff pot}, we give general formulae regarding the effective potential. We start with the BRST-quantization of our toy model and find the mass spectrum and free propagators in the Feynman gauge. Then, we obtain a formal expression for the one-loop effective action and reduce the problem to a finding of an analytic expression for $\Sp\ln(\nabla^2+m^2)$ on the Schwarzschild background. Sec. \ref{oneloop} is devoted to this problem, where it is solved in the quasiclassical approximation. Of course, we need some regularization prescription there to remove divergencies. This is done by introducing a certain cutoff parameter. It turns out that the obtained expression for the one-loop correction is immediately generalized to an arbitrary static, spherically symmetric background metric. Also, in this section we investigate how the result depends on the regularization scheme applied. The form of a mass dependent part of the effective potential proves to be independent of the regularization scheme. Moreover, under the assumption that the conformal symmetry takes place in the massless case the form of the whole effective potential is fixed unambiguously. In Sec. \ref{analysis}, we describe the renormalization procedure and provide a brief analysis of the physical implications following from the obtained one-loop effective potential. Possible scenarios of the physics near a black hole are established. In Appendix, we give asymptotic expansions of the integrals encountered in calculating the first quantum correction in Sec. \ref{oneloop}.

We use the following notation and conventions. The system of units is chosen so that the velocity of light $c$, the Planck constant $\hbar$, and the Schwarzschild radius $r_g$ of a black hole are unities. In Sec. \ref{analysis}, where we analyze the physical effects near a black hole, we restore the ordinary length units. Greek letters denote space-time indices. Sometimes we shall use boldface characters to denote the spatial part of coordinates. Also, we use the mostly minus signature of the space-time metric $g_{\mu\nu}$.

\section{Effective potential: general formulae}\label{eff pot}

Consider the model of the $SO(N)$ scalar multiplet $\phi^n(x)$ in the fundamental representation, $n=\overline{1,N}$, minimally coupled to the gauge fields $A_\mu(x)=A_\mu^at_{an}^m$, $a=\overline{1,N(N-1)/2}$, on the Schwarzschild background $g_{\mu\nu}(x)$.  The action functional of the model reads
\begin{equation}\label{action}
    S[\phi,A_\mu]=\int d^4x\sqrt{|g|}\left[\frac12(D_\mu\phi^n)^2-V_{0}(\phi)-\frac14 G_{\mu\nu}^aG^{\mu\nu}_a\right],
\end{equation}
where $V_{0}(\phi)=\La+\mu^2\phi^2/2+\la(\phi^2)^2/4$,
\begin{equation}
    D_\mu\phi=\partial_\mu\phi+ieA_\mu\phi, \qquad G_{\mu\nu}^a=\partial_{[\mu}A_{\nu]}^a+ef^a_{bc}A^b_\mu A^c_\nu,
\end{equation}
and $\La$, $\mu$, $\la$, and $e$ are the coupling constants, $f^a_{bc}$ are the structure constants of the Lie algebra $so(N)$ with generators $t_a$:
\begin{equation}
    [t_a,t_b]=if_{ab}^ct_c,\qquad t^+_a=t_a,\qquad t^T_a=-t_a,
\end{equation}
where transposition is understood with respect to the Euclidean metric preserved by the action of the $SO(N)$ group. The Latin indices $a,b,c\ldots$ are risen and lowered by the Killing metric and the Greek indices $\mu,\nu\ldots$ are risen and lowered by the space-time metric. Henceforth the matrix notation is implied and the indices are suppressed when it does not lead to confusion. The $SO(N)$ symmetry of the model is spontaneously broken by the Higgs mechanism down to $SO(N-1)$. The Higgs field can be decomposed as
\begin{equation}
    \phi(x)=\eta(x)+\chi(x),
\end{equation}
where the vector $\eta(x)$ is its vacuum expectation value.  It provides a minimum to the classical potential $V_{0}(\phi)$ in the tree approximation.

As the model \eqref{action} possesses the gauge symmetries, we need to use the BRST-quantization procedure (see, e.g., \cite{HeTe,FaddSlav,Wein}) in order to pass to the quantum theory consistently. The quantization procedure is quite standard, but it is usually given for a constant background. Therefore, we trace some its basic steps for our case.

It is convenient to change the coordinates so that $g^{00}=1$ and $g^{0i}=0$, and return to the original ones in the total action with ghosts. Introducing the momenta
\begin{equation}
\begin{gathered}
    \Pi^i_a=-G^{0i}_a,\qquad\pi_n=D^0\phi_n,\\
    \{\phi_n(t,\spx),\pi^m(t,\spy)\}=\frac{\de^m_n(\spx-\spy)}{\sqrt{|g|}},\qquad\{A_i^a(t,\spx),\Pi^j_b(t,\spy)\}=\frac{\de^a_b\de^i_j(\spx-\spy)}{\sqrt{|g|}},
\end{gathered}
\end{equation}
where curly brackets denotes the Poisson brackets, we arrive at the Hamiltonian action
\begin{multline}\label{action ham}
    S_H[\phi,\pi,A_\mu,\Pi^i]=\int d^4x\sqrt{|g|}\Bigl[\Pi_a^i\dot{A}^a_i+\pi\dot{\phi}+\frac12\Pi^a_i\Pi^i_a-\frac14 G^a_{ij}G^{ij}_a\\
    +A^a_0(\bar{\nabla}_i\Pi^i_a-ie\pi t_a\phi)-\frac12\pi^2+\frac12D_i\phi D^i\phi-V_{0}(\phi) \Bigr],
\end{multline}
where the overdots denote the derivatives with respect to time and
\begin{equation}
    \bar{\nabla}_i\Pi^i_a=\nabla_i\Pi^i_a+e\Pi^i_bf^b_{ac}A^c_i=|g|^{-1/2}\partial_i(|g|^{1/2}\Pi^i_a)+e\Pi^i_bf^b_{ac}A^c_i.
\end{equation}
As usual for the relativistic invariant models, $A^a_0$ are the Lagrange multipliers to the first class constraints
\begin{equation}
    T_a:=\nabla_i\Pi^i_a-ie\pi t_a\phi,\qquad\{T_a(t,\spx),T_b(t,\spy)\}=ef^c_{ab}T_c(t,\spx)\frac{\de(\spx-\spy)}{\sqrt{|g|}}.
\end{equation}
Therefore, the BRST-charge is
\begin{equation}
    \Omega=\int d\spx\sqrt{|g|}\left[c^aT_a+\bar{c}^a\Pi_a^0-\frac12c^ac^bf_{ab}^cP_c\right],
\end{equation}
where we introduce the scalar fermionic ghost fields $(c^a,P_a)$ and $(\bar{c^a},\bar{P_a})$ with the Poisson brackets
\begin{equation}
    \{c^a(t,\spx),P_b(t,\spy)\}=\{\bar{c}^a(t,\spx),\bar{P}_b(t,\spy)\}=\frac{\de^a_b(\spx-\spy)}{\sqrt{|g|}},\qquad\gh
    c=\gh\bar{c}=-\gh P=-\gh\bar{P}=1.
\end{equation}
Other fields are graded to have zero ghost number. We choose the gauge fixing fermion $\psi$ so it results in the Feynman gauge
\begin{equation}
    \psi=\int d\spx\sqrt{|g|}\left[-P_aA^a_0-\bar{P}_a\left(\nabla_i A^{ai}+ie\chi t^a\eta+\frac12\Pi_0^a\right)\right],\qquad\gh\psi=-1.
\end{equation}
Adding $-\{\Omega,\psi\}$ to the Hamiltonian action \eqref{action ham} taken on the constraint surface, we obtain the total Hamiltonian action with the Lagrangian
\begin{multline}
    L^H_{tot}=\sqrt{|g|}\Bigl[\Pi_a^i\dot{A}^a_i+\pi\dot{\phi}+\Pi^0_a\left(\nabla_\mu A^{a\mu}+ie\chi t^a\eta+\frac12\Pi^a_0\right)+\frac12\Pi^a_i\Pi^i_a-\frac14 G^a_{ij}G^{ij}_a
    +A^a_0(\bar{\nabla}_i\Pi^i_a-ie\pi t_a\phi)\\-\frac12\pi^2+\frac12D_i\phi D^i\phi-V_{0}(\phi)+\bar{c}^a(\dot{\bar{P}}_a-P_a)+c^a\dot{P}_a-\partial^ic^a\partial_i\bar{P}_a+ec^af^b_{ac}(A^c_i\partial^i\bar{P}_b+A^c_0P_b)-e^2c^a\phi t_a^Tt_b\eta\bar{P}^b\Bigr].
\end{multline}
Substituting this action into the functional integral and integrating over the fields $\bar{c}$, $P$, $\Pi$ and $\pi$, we have (cf. \cite{Wein})
\begin{multline}\label{action tot}
    L_{tot}=\sqrt{|g|}\Bigl[c^a(\nabla^2-e^2\eta t_a^Tt_b\eta)\bar{P}^b+\frac12A_\mu^a(\nabla^2-e^2\eta t^T_at_b\eta)A^{b\mu}-\frac12\chi(\nabla^2-e^2t_a\eta\eta t_a^T+V''_{0}(\eta))\chi\\
    -V_0(\eta+\chi)+\frac12\chi V''_0(\eta)\chi+\partial^\mu\chi\partial_\mu\eta+\frac12\partial_\mu\eta\partial^\mu\eta-\frac{ie}2(\partial_\mu\eta A^\mu\eta-\eta A^\mu\partial_\mu\eta)-ie\partial_\mu\chi A^\mu\chi
    -e^2\chi A^T_\mu A^\mu\eta\\
    -\frac{e^2}2\chi A_\mu^TA^\mu\chi
    +c^a(f^b_{ac}A^c_\mu\partial^\mu-\chi t_a^Tt_b\eta)\bar{P}_b-\frac{e}2\partial_{[\mu}A^a_{\nu]}f^a_{bc}A^{b\mu}A^{c\nu}-\frac{e^2}4(f^a_{bc}A^{b\mu}A^{c\nu})^2\Bigr],
\end{multline}
where $V''_0:=\partial^2V_0/\partial\phi^n\partial\phi^m$ and we take into account that the space-time is Ricci flat: $R_{\mu\nu}=0$. By the same reason we did not add the nonminimal renormalizable coupling $\xi R\phi^2$ to the action \eqref{action}. Such a term is exploited in some inflationary models \cite{BezShap}. Even if the gravitational field does not satisfy the vacuum Einstein equations these terms will be irrelevant for our subsequent analysis. We shall make all calculations in the quasiclassical approximation (see below) and these contributions can be neglected. Upon integration over momenta a local contribution to the path-integral measure appears (see, for example, \cite{LyakhovichPhD}), but as we shall see such contributions are irrelevant for our calculations.

The first line in \eqref{action tot} contains the terms quadratic in the fields and it is only needed to find the first quantum correction to the effective action. It is not difficult to deduce the mass spectrum of particles. Bearing in mind that for our model
\begin{equation}
    V''_0(\eta)t_a\eta=0,\qquad\eta t_a^Tt^c\eta\eta t_c^Tt^b\eta=-\eta^2\eta t_a^Tt^b\eta,\qquad\eta t_a^Tt^a\eta=\eta^2(1-N),
\end{equation}
we conclude that there are $2(N-1)$ ghosts, $4(N-1)$ vector bosons, and $(N-1)$ Goldstone bosons with the squared mass
\begin{equation}\label{W mass}
    m^2=e^2\eta^2,
\end{equation}
the one Higgs boson with the squared mass
\begin{equation}\label{Higgs mass}
    m_H^2=\mu^2+3\la\eta^2,
\end{equation}
and the massless fields: $(N-1)(N-2)$ ghosts and $2(N-1)(N-2)$ vector bosons. Under the number of particles we mean the number of their states.

The value of the effective action at $\phi=\eta$ when the other fields are integrated out is given by general formula (see, e.g., \cite{Wein})
\begin{equation}\label{effective action gen}
    i\Ga[\eta]=\int D\chi DA DcD\bar{P} e^{iS_{tot}[\eta+\chi,A,c,\bar{P}]},
\end{equation}
where the integral assumes a summation over the vacuum, one-particle irreducible, connected diagrams only. In view of the last remark, the linear in the fields $\chi$ and $A_\mu$ terms entering \eqref{action tot} can be omitted. The one-loop contribution to the effective action is formally given by
\begin{multline}\label{action oneloop gen}
    i\Ga_1[\eta]=\frac{N-1}2\Sp\ln(\nabla^2+m^2)-\frac{N-1}2\Sp\ln(\nabla^{2\mu}_\nu+m^2\de^\mu_\nu)-\frac12\Sp\ln(\nabla^2+m^2_H)\\
    +\frac{(N-1)(N-2)}2\Sp\ln\nabla^2-\frac{(N-1)(N-2)}4\Sp\ln\nabla^{2\mu}_\nu+\ldots
\end{multline}
where the dots denote higher terms of the loop expansion, which are small as long as the renormalized coupling constants $e$ and $\la$ are small. Also we omit the contributions proportional to the trace of the identity operator. The Feynman propagators appearing in the logarithms \eqref{action oneloop gen} are specified in the standard way \cite{DeWittDTGF} by the $i\epsilon$-prescription $m^2\rightarrow m^2-i\epsilon$. The operators $\nabla^2$ and $\nabla^{2\mu}_\nu$ are defined as follows
\begin{equation}\label{Laplace operators}
\begin{gathered}
    \nabla^2=|g|^{-1/2}\partial_\mu|g|^{1/2}g^{\mu\nu}\partial_\nu,\\
    \nabla^{2\mu}_\nu=\nabla^2\de^\mu_\nu+g^{\la\rho}[\partial_\la(\Ga^\mu_{\rho\nu})+2\Ga^\mu_{\rho\nu}\partial_\la+\Ga^\mu_{[\la\s}\Ga^\s_{\nu]\rho}]=\nabla^2\de^\mu_\nu+g^{\la\rho}[\partial_\nu(\Ga^\mu_{\la\rho})+2\Ga^\mu_{\rho\nu}\partial_\la],
\end{gathered}
\end{equation}
where the square brackets enclosing indices mean an antisymmetrization without one half. In the last equality, we used the fact that the Schwarzschild metric is the solution of the vacuum Einstein equations. Formula \eqref{action oneloop gen} is valid up to  contributions from the path-integral measure. It can be shown \cite{DeWittDTGF} that the path-integral measure respecting unitarity of the $S$-matrix gives a one-loop correction to the effective action of the same form as \eqref{action oneloop gen}, but with an opposite sign and the Feynman propagators replaced by the advanced Green functions. A general scheme how to define and calculate the $S$-matrix on a curved background can be found, for example, in \cite{BuchGit}.

In the next section, we shall calculate the traces entering \eqref{action oneloop gen} and the analogous expression coming from the measure in the quasiclassical (short-wave) approximation. In this approximation, the derivatives in \eqref{Laplace operators} acting on the right are of the order of unity and the connections are of the order of $\hbar$ (see for details \cite{MaslFed,BBT}). Hence
\begin{equation}
    \nabla^{2\mu}_\nu=\nabla^2\de^\mu_\nu+O(\hbar).
\end{equation}
This approximation is adequate for the modes with the wave-lengths much smaller than the gravitational radius $r_g$. It can be interpreted as we effectively integrate over the fast modes only in the functional integral \eqref{effective action gen}. The correction to the effective action \eqref{action oneloop gen} due to the slow modes with the characteristic wave-lengths larger than $r_g$ is well approximated by the total action \eqref{action tot} with renormalized fields and coupling constants. The ghost fields can be excluded since they do not influence the dynamics of the mean fields $\chi$ and $A_\mu$. Thus, the one-loop correction to the effective action in the quasiclassical limit reads
\begin{equation}
  i\Ga_1[\eta]\approx-\frac32(N-1)\Sp\ln\frac{(\nabla^2+m^2)_F}{(\nabla^2+m^2)_+}-\frac12\Sp\ln\frac{(\nabla^2+m^2_H)_F}{(\nabla^2+m^2_H)_+}
    -\frac{(N-1)(N-2)}2\Sp\ln\frac{(\nabla^2)_F}{(\nabla^2)_+},
\end{equation}
where the subscripts indicate what the Green function should be taken. The main problem facing us is to derive analytical expressions for $\Sp\ln(\nabla^2+m^2)_{F,+}$.

\section{One-loop correction}\label{oneloop}

In this section, we find the one-loop correction to the effective potential given by quantum fluctuations of one scalar bosonic field in the quasiclassical approximation on the Schwarzschild background. It turns out, the expression we shall obtain is valid for an arbitrary static, spherically symmetric background.

To this end, we write
\begin{equation}\label{spur log}
%\begin{split}
    -\frac12\Sp\ln(\nabla^2+m^2)\approx-\frac12\int d^4x\sqrt{|g|}\sum\limits_\xi|\psi_\xi(x)|^2\ln(-\e(\xi)+m^2(x)),\qquad
    \int d^4x\sqrt{|g|}|\psi_\xi(x)|^2=1,
%\end{split}
\end{equation}
where $\psi_\xi(x)$ are the eigenfunctions of the wave operator $\nabla^2$ and $-\e(\xi)$ is its spectrum, $\xi$ specifies the quantum numbers. The approximate equality appears since we assume that $m^2$ depends on the space-time point and so Eq. \eqref{spur log} is valid in the leading order in $\hbar$ only. The correction to this equality is proportional to a commutator of the wave operator with $m^2(x)$ and we neglect it. Inasmuch as we work in the Schwarzschild coordinates, the interval is
\begin{equation}
    ds^2=(1-r^{-1})dt^2-(1-r^{-1})^{-1}dr^2-r^2(d\theta^2+\sin^2\theta d\vf^2).
\end{equation}
Besides, we assume for a while that $\psi_\xi(x)$ vanishes at the big spatial sphere surrounding the black hole and is $T$-periodic with respect to the time variable. Hence, we have to solve the  equation
\begin{equation}\label{wave eq}
    \left[(1-r^{-1})^{-1}\partial_t^2-\frac1{r^2}\partial_r(r^2-r)\partial_r-\frac{1}{r^2\sin^2\theta}(\sin\theta\partial_\theta\sin\theta\partial_\theta+\partial^2_\vf)\right]\psi=-\e\psi,
\end{equation}
with the given boundary conditions. Separating variables
\begin{equation}
    \psi_{nklm}=e^{iEt}Y_{lm}(\theta,\vf)R_{nkl}(r),\qquad E=2\pi k/T,
\end{equation}
where $Y_{lm}(\theta,\vf)$ are the spherical harmonic functions, we reduce Eq. \eqref{wave eq} to the radial one
\begin{equation}\label{radial eq}
    \left[-\frac1{r^2}\partial_r(r^2-r)\partial_r-(1-r^{-1})^{-1}E^2+\frac{l(l+1)}{r^2}\right]R_{nkl}=-\e_{nkl} R_{nkl}.
\end{equation}
It can be solved in terms of the Heun functions (see, for example, \cite{Valent}), but the properties of these functions are not well investigated. The spectral problem \eqref{radial eq} was studied in many respects and the approximate solutions to it were found under certain assumptions (see, e.g., \cite{DeWittDTGF,spectrum,FrolNov}). However, keeping in mind that we are interested not in the spectrum itself, but in \eqref{spur log}, we shall try for another strategy and seek for a formal quasiclassical solution of \eqref{radial eq} at arbitrary values of the energy and angular momentum. We shall not write the Planck constant explicitly as it is restored in an obvious manner.

Making the standard substitution,
\begin{equation}\label{radial quasi}
    R(r)=f(r)e^{iS(r)},
\end{equation}
in \eqref{radial eq} we arrive in the two leading orders in $\hbar$ at
\begin{equation}
    S'=\pm(1-w)^{-1}\sqrt{E^2-(1-w)(\e+L^2w^2)},\qquad f=\frac{c}{\sqrt{|(r^2-r)S'|}},
\end{equation}
where $w:=r^{-1}$, $L^2:=l(l+1)$, and $c$ is some constant. The turning points are at the the vanishing radicand
\begin{equation}\label{turning p}
    E^2-(1-w)(\e_{nkl}+L^2w^2)=0,
\end{equation}
or at the boundary sphere or at the origin of coordinates. The spectrum is determined by the requirement that \eqref{radial quasi} is a single-valued function
\begin{equation}\label{BS quantization}
    \int dr(1-w)^{-1}\sqrt{E^2-(1-w)(\e_{nkl}+L^2w^2)}=\pi n+\ga,\quad n\in \mathbb{Z},
\end{equation}
where the integral is taken between two turning points and $\ga$ depends on whether the particle strikes the boundary sphere or reaches the origin or not. The values $\pi/4$, $\pi/2$, and $3\pi/4$ of $\ga$ correspond to the cases: (i) the particle moves between the origin and the turning point \eqref{turning p}, (ii) moves between two turning points \eqref{turning p} or between the origin and the boundary sphere, and (iii) moves between the turning point \eqref{turning p} and the boundary sphere. It can be proved \cite{BBT,MaslFed} that the spectrum derived in this way differs from the exact one by the terms of the order of $\hbar^2$. The normalization condition \eqref{spur log} becomes
\begin{equation}\label{normalization}
    T|c|^2\int_{r_1}^{r_2} \frac{dr}{\sqrt{E^2-(1-w)(\e+L^2w^2)}}=1,
\end{equation}
where $r_1$ and $r_2$ are the turning points and we use the standard normalization of the spherical harmonics:
\begin{equation}
    \int d\vf d\theta\sin\theta Y^*_{lm}(\theta,\vf)Y_{l'm'}(\theta,\vf)=\de_{ll'}\de_{mm'}.
\end{equation}

Recall that the spherical harmonic functions possesses the property
\begin{equation}
    \sum\limits_{m=-l}^l|Y_{lm}(\theta,\vf)|^2=\frac{2l+1}{4\pi}.
\end{equation}
Then the integrand of \eqref{spur log} can be written as
\begin{equation}\label{integrand}
    I:=-\frac12\sum\limits_\xi|\psi_\xi(x)|^2\ln(-\e(\xi)+m^2)=-\frac1{8\pi}\sum\limits_{l=0}^\infty(2l+1)\sum\limits_{n,k=-\infty}^\infty\frac{|c_{nkl}|^2\ln(-\e_{nkl}+m^2)}{r^2\sqrt{E^2-(1-w)(\e_{nkl}+L^2w^2)}}.
\end{equation}
In the quasiclassical approximation and in the limit $T\rightarrow\infty$ whereas the radius of the boundary sphere tends to infinity, it is admissible to pass from a summation over the quantum numbers to an integration over them. The Bohr-Sommerfeld quantization condition \eqref{BS quantization} and the normalization \eqref{normalization} entails
\begin{equation}\label{dn}
    dn=-\frac{d\e}{2\pi}\int_{r_1}^{r_2} \frac{dr}{\sqrt{E^2-(1-w)(\e+L^2w^2)}}=-\frac{d\e}{2\pi T|c|^2}.
\end{equation}
The Jacobian is nonzero everywhere. Thus, we obtain for the one-loop correction \eqref{integrand}
\begin{equation}\label{I_II}
  I=-\frac1{8\pi}\int_{-\infty}^\infty\frac{dE}{2\pi}\int_0^\infty \frac{dL^2}{r^2}\int_{-\infty}^\infty \frac{d\e}{2\pi}\frac{\theta(E^2-(1-w)(\e+L^2w^2))}{\sqrt{E^2-(1-w)(\e+L^2w^2)}}\ln(-\e+m^2),
\end{equation}
where $\theta(x)$ is the Heaviside step function. Tracing all the steps of our derivation starting from \eqref{wave eq}, it is not difficult to see that for an arbitrary static, spherically symmetric metric in the adapted coordinates we arrive at
\begin{multline}\label{I_II_gen}
    I=\frac1{8\pi}\int_{-\infty}^\infty\frac{dE}{2\pi}\int_0^\infty \frac{2dL^2}{\xi^2_L}\int_{-\infty}^\infty \frac{d\e}{2\pi}\frac{\theta(E^2-\xi^2_t(\e-2L^2/\xi^2_L))}{\sqrt{E^2-\xi^2_t(\e-2L^2/\xi^2_L)}}\ln(-\e+m^2)\\
    =-\frac1{8\pi}\int_{-\infty}^\infty\frac{dE}{2\pi}\int_0^\infty \frac{dL^2}{r^2}\int_{-\infty}^\infty \frac{d\e}{2\pi}\frac{\theta(E^2-g_{00}(\e+L^2w^2))}{\sqrt{E^2-g_{00}(\e+L^2w^2)}}\ln(-\e+m^2),
\end{multline}
where we assume that $g_{00}g_{rr}\leq0$, and $\xi$'s denote the Killing vectors
\begin{equation}
    \xi_t:=\partial_t,\qquad\xi_L^2:=g_{\mu\nu}(\xi_x^\mu\xi_x^\nu+\xi_y^\mu\xi_y^\nu+\xi_z^\mu\xi_z^\nu)=-2r^2,
\end{equation}
with $\xi_i$ being the generators of rotations. In passing, we cast the integral $I$ into an explicit generally covariant form. According to the $i\epsilon$-prescription for the Feynman propagator, the logarithm should be written as
\begin{equation}
    \ln(-\e+m^2)_F=\ln|-\e+m^2|-i\pi\theta(\e-m^2).
\end{equation}
As for the advanced Green function we specify it by adding $i\epsilon$ to the energy $E$. This is a correct definition of the advanced Green function, at least, for the space-time points above the horizon of a black hole. It is that region which we shall be interested in. Hence, for the advanced Green function the logarithm should be replaced by
\begin{equation}\label{log adv}
    \ln(-\e+m^2-i\epsilon g^{00}E)=\ln|-\e+m^2|-i\pi\sign(g_{00}E)\theta(\e-m^2),
\end{equation}
in the quasiclassical approximation.

The obtained integral \eqref{I_II_gen} has ultraviolet divergencies. Since this is a multiple integral it should be regularized with some care. In contrast to the case of a flat space-time, there are not continuous symmetries like the Lorentz symmetry which we should preserve. The only residual symmetry is the discrete $T$-symmetry $E\rightarrow-E$. Therefore, we regularize the integral cutting the range of integration over the respective variables preserving the time reversion symmetry. At that, we apply this regularization prescription to the initial integral \eqref{I_II} or more rigorously to the initial sum \eqref{integrand} and assume that the regularization parameters do not depend on $m^2$ and $r$. In other words, we can find the derivatives of the effective  Lagrangian \eqref{integrand} with respect to $m^2$ and $r$ by a mere differentiation of the expression under sums\footnote{Such a trick is a standard tool of the regularization on a flat space-time. Unfortunately, it is unapplicable in our case since the integral remains divergent after differentiation and we lose the information about the terms of the effective potential at lower powers of $m^2$. They are not constant as in the flat space-time.}.
Then, making changes of variables, which do not depend on $m^2$ and $r$, we reduce this integral to the integral which is divergent in one variable only. That is, the ``singular'' boundary of the integration domain is retracted to a point.  The regularization procedure of singular integrals in one variable is well developed (see, e.g., \cite{GSh}). We simply cut off the divergence of this integral. The emergent regularization parameter is some function of the initial regularization parameters which does not depend on $m^2$ and $r$. An exact form of this function is irrelevant for us. All we need to know about it is that it tends to the regularization removal limit as the initial regularization parameters do. Notice that the change of variable \eqref{dn} is independent of $m^2$ and $r$.

Consider the contribution \eqref{I_II_gen} to the effective Lagrangian at the point above the horizon of the black hole ($w<1$). The integral over the squared angular momentum $L^2$ is easily taken with the result
\begin{equation}\label{I_int}
    I=-\int\frac{dEd\e}{(2\pi)^2}\theta(E^2-y\e)\frac{\sqrt{E^2-y\e}}{4\pi y}\ln(-\e+m^2)_F,
\end{equation}
where $y\equiv g_{00}=\xi^2_t$. Despite the real part of this integral does not contribute to the effective Lagrangian, because it is canceled out by the same contribution from the advanced Green function, we investigate its asymptotic expansion as well. In order to take the obtained integral, it is useful to pass into the polar system of coordinates
\begin{equation}
    \e=\rho\cos\vf,\qquad E^2=\rho\sin\vf.
\end{equation}
Then the integral over the angular variable $\vf$ converges, whereas the integral over $\rho$ diverges quadratically. We regularize this divergence introducing the cutoff parameter $\De$
\begin{equation}
    I=-\int_{\arctan y}^\pi\frac{d\vf}{16\pi^3}\frac{\sqrt{1-y\cot\vf}}y\int_0^\De d\rho\rho[\ln|-\rho\cos\vf+m^2|-i\pi\theta(\rho\cos\vf-m^2)].
\end{equation}
After a suitable change of variables, we come to
\begin{multline}\label{reI}
    \re I=-\int_{-1}^{(1+y^2)^{-1/2}}\frac{dx}{32\pi^3y}\frac{(\sqrt{1-x^2}-xy)^{1/2}}{(1-x^2)^{3/4}}\biggl[\De^2\ln\De-\frac{\De^2}2+\De^2\ln\left|x-\frac{m^2}{\De}\right|\\
    -\frac{m^4}{x^2}\left(\ln\left|1-\frac{x\De}{m^2}\right|+\frac{x\De}{m^2}\right)\biggr],
\end{multline}
and
\begin{equation}\label{imI}
    \im I=\frac{m^4}{32\pi^2y}\int_{(\De^2/m^4-1)^{-1/2}}^{y^{-1}}dx\frac{\sqrt{1-xy}}{1+x^2}\left(\frac{\De^2}{m^4}-1-\frac1{x^2}\right).
\end{equation}

To obtain an asymptotic expansion of the real part of \eqref{reI}, in terms of the regularization parameter we use the formulae \eqref{ap:L} and \eqref{ap:M} given in Appendix. Then its singular and finite parts look like
\begin{multline}\label{reI1}
    -\re I=\frac{1}{32\pi^3y}\biggl\{\frac{\De^2}2\ln\frac{\De^2}eI_0(y)+\De^2\tilde{I}_0(y)-2\De m^2\left[I_1(y)-\ln\sqrt{1+y^2}\right]\\
    -\frac{m^4}2\biggl[I_2(y)+2\tilde{I}_2(y)-3-3\sqrt{1+y^2}-\frac{y}2\left(\ln^2\sqrt{1+y^2}-\ln\sqrt{1+y^2}-\pi^2\right)+\sqrt{1+y^2}\ln(1+y^2)\\
    +2\ln\frac{m^2}\De\left(1+\sqrt{1+y^2}-\frac{y}2\ln\sqrt{1+y^2}-I_2(y)\right)\biggr] \biggr\},
\end{multline}
where
\begin{equation}\label{Is}
\begin{aligned}
    I_0(y)&:=\int dxZ(x)=\frac{\pi y}{\sqrt{2\sqrt{1+y^2}-2}},\\
    \tilde{I}_0(y)&:=\int dxZ(x)\ln|x|=\frac{y}2\int_{-\infty}^1dx\frac{\sqrt{1-x}}{x^2+y^2}\ln\frac{x^2}{x^2+y^2},\\
    I_1(y)&:=\int dx\frac{Z_1(x)}x=y\int_{-\infty}^1\frac{dx}{x(x^2+y^2)}\left[\sqrt{(1-x)(x^2+y^2)}-y\right],\\
    I_2(y)&:=\int dx\frac{Z_2(x)}{x^2}=1+\sqrt{1+y^2}-\frac{y}2\ln\sqrt{1+y^2},\\
    \tilde{I}_2(y)&:=\int dx\frac{Z_2(x)}{x^2}\ln|x|=\frac{y}2\int_{-\infty}^1\frac{dx}{x^2}\left[\sqrt{1-x}-\frac{y}{\sqrt{x^2+y^2}}+\frac{y^2}2\frac{x}{x^2+y^2}\right]\ln\frac{x^2}{x^2+y^2}.
\end{aligned}
\end{equation}
All the integrals on the left hand side are taken over $x\in[-1,(1+y^2)^{-1/2}]$ and
\begin{equation}
    Z(x)=\frac{(\sqrt{1-x^2}-xy)^{1/2}}{(1-x^2)^{3/4}}=1-\frac{xy}2+O(x^2).
\end{equation}
The integrals $\tilde{I}_0$ and $\tilde{I}_2$ are reduced to dilogarithms, while $I_1$ is an elliptic integral. We see that the term at $m^4\ln m^2$ in Eq. \eqref{reI1} vanishes. Introducing new notations for the infinite constants entering the real part  \eqref{reI1}, we can write
\begin{multline}\label{reI 1}
    -\re I=B_0\frac{I_0+\be\tilde{I}_0}{y}+B_1 \frac{m^2}y\left[I_1(y)-\ln\sqrt{1+y^2}\right]\\
    -\frac{m^4}{32\pi^3y}\left[\tilde{I}_2(y)-1+\sqrt{1+y^2}\left(\ln\sqrt{1+y^2}-1\right)-\frac{y}4\left(\ln^2\sqrt{1+y^2}-\pi^2\right)\right]\\
    =:B_0g_0(y)+B_1 m^2g_1(y)
    -\frac{m^4}{32\pi^3} g_2(y),
\end{multline}
where $\beta$ is some dimensionless constant.

\begin{figure}[t]
\centering%
\includegraphics*[width=7cm]{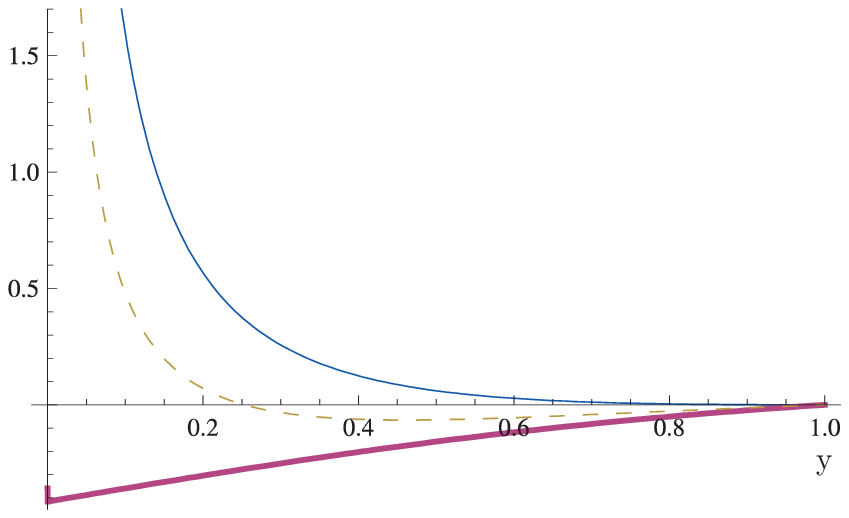}\;
\includegraphics*[width=7cm]{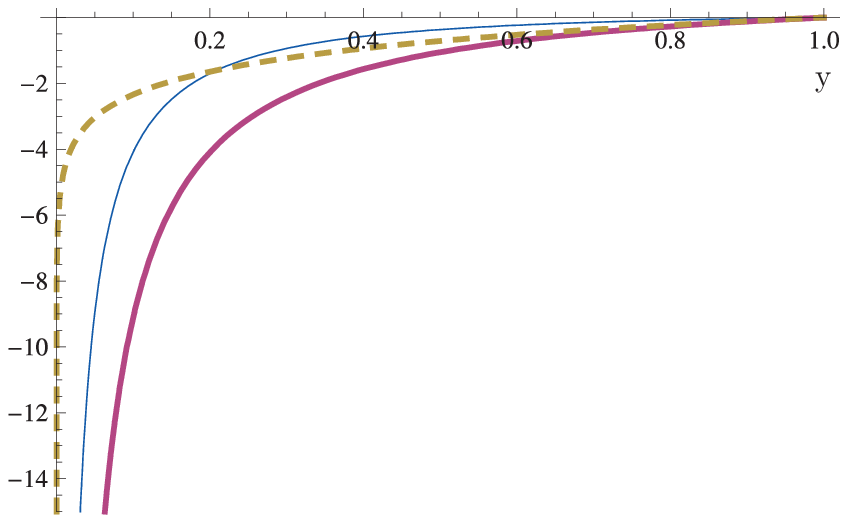}
\caption{{\footnotesize At the left: the function $\bar{g}_0(y)$ at $\be=1/\ln2$ (thick line), at $\be=1.4$ (dashed line), and at $\be=1.34$. At the right: the function $\bar{g}_2(y)/\pi$ (thick line), the function $\bar{g}_1(y)$ (dashed line), and $-\overline{(R(y)/\pi y)}$. Bars over the functions mean that we subtract from them their values at $y=1$.}}\label{plot of functions}
\end{figure}

The integral \eqref{imI} is easily expanded in $(\De^2/m^4-1)^{-1/2}$. So, up to the terms vanishing at $\De\rightarrow\infty$, the one-loop contribution to the effective Lagrangian reads
\begin{multline}\label{Spln calc}
    L^{eff}_1=\im I=\frac1{32\pi^2}\left\{\De^2\frac{R(y)}{y}-2\De\frac{m^2}y+\frac{m^4}2\left(\frac32+\ln(4\De)\right)-\frac{m^4}2\ln(ym^2)\right\}\\
    =A_0\frac{R(y)}{y}+A_1\frac{m^2}y+A_2m^4-\frac{m^4}{64\pi^2}\ln(ym^2),
\end{multline}
where
\begin{multline}\label{R}
    R:=y\int_0^1dx\frac{\sqrt{1-x}}{x^2+y^2}=\sqrt{\frac12\sqrt{1+y^2}+\frac12}\arccot\sqrt{\frac12\sqrt{1+y^2}-\frac12}\\
    -\sqrt{\frac12\sqrt{1+y^2}-\frac12}\arccoth\sqrt{\frac12\sqrt{1+y^2}+\frac12}.
\end{multline}
The analogous contribution \eqref{log adv} from the advanced Green function is zero. Recall that the obtained expression is valid for the points above the horizon of any static, spherically symmetric metric. In spite of we use the system of units in which a mass is measured in $\hbar r_g^{-1}c^{-1}$, formula \eqref{Spln calc} holds as is in the system of units $\hbar=c=1$ with the standard mass units (say, GeV/$c^2$). The undetermined dimensional constants are only redefined and $y=1-r_g/r$. The plots of the functions $g_i(y)$ and $R(y)/y$ are presented at Fig. \ref{plot of functions}. Asymptotics of these functions are as follows: at the horizon
\begin{equation}\label{fg expans hor}
\begin{aligned}
    R/y&=\frac{\pi}2y^{-1}+\frac12\left(\ln\frac{y}4-1\right)+\frac\pi{16}y+o(y),\\
    g_0&=\pi(1-\be\ln2)y^{-1}+\frac\pi8(1-\be\ln2+\be)y+o(y),\\
    g_1&=\ln y-1-3\ln2+o(1),\\
    g_2&=-\pi y^{-1}+o(y^{-1}),
\end{aligned}
\end{equation}
while at the spatial infinity
\begin{equation}\label{fg expans}
\begin{aligned}
    R/y&=0.561-0.971(y-1)+1.23(y-1)^2+O((y-1)^3),\\
    g_0&=3.45-2.10\be-(2.95-2.19\be)(y-1)+(3.04-2.22\be)(y-1)^2+O((y-1)^3),\\
    g_1&=-3.04+1.02(y-1)-0.534(y-1)^2+O((y-1)^3),\\
    g_2&=-0.313+3.45(y-1)-3.20(y-1)^2+O((y-1)^3).
\end{aligned}
\end{equation}
We provide the approximate numerical values for the expansion coefficients at the spatial infinity. Their exact expressions through the integrals or arccotangents of huge arguments can be read off from Eqs. \eqref{Is}, \eqref{R}.

Let us study how the obtained asymptotics depend on a shape of the integration domain in the space of the quantum numbers. The asymptotic expansion of the imaginary part \eqref{imI} for an arbitrary star-shaped region of integration with the boundary $\De(x)$, $\De(x)\rightarrow\infty$, takes the form
\begin{multline}\label{Spln calc 1}
    L_1^{eff}=\int_0^{y^{-1}}\frac{dx}{32\pi^2 y}\frac{\sqrt{1-xy}}{1+x^2}\De^2(x)-\frac{\De(0)}{16\pi^2}\frac{m^2}y\\
    +\frac{m^4}{32\pi^2}\left[\frac12\left(\frac32+\ln[4\De(0)]\right)-y^{-1}\frac{\De'(0)}{\De(0)}\right]-\frac{m^4}{64\pi^2}\ln(ym^2),
\end{multline}
where $x=0$ corresponds to massless modes. So, the part of the one-loop contribution to the effective Lagrangian depending on a mass is completely determined by three constants irrespective of an integration domain, which, of course, should coincide with the whole space of quantum numbers in the regularization removal limit. The massless contribution depends considerably on the shape of the integration region. We can fix the form of $\De(x)$ unambiguously if demand that in the massless case the effective action possesses a conformal invariance on the given background. From Eq. \eqref{I_II_gen} and \eqref{I_int} it is not hard to see that a stretching of the metric by a function $\la(r)$ is equivalent to a stretching of $y$ by this function. Therefore, if we take $\De^2(x)=(1+x^2)\De^2(0)$, the one-loop contribution to the effective action,
\begin{equation}\label{Spln calc new}
    L_1^{eff}=A_0y^{-2}+A_1m^2y^{-1}+A_2m^4+\tilde{A}_2m^4y^{-1}-\frac{m^4}{64\pi^2}\ln(ym^2),
\end{equation}
will be conformal invariant in the massless case. The factor $\la^{-2}$ is canceled by the factor coming from the measure. The choice $\De^2(x)=(1+x^2)\De^2(0)$ corresponds to a usual energy cutoff, i.e., all the modes with the energies higher than $\De^{1/2}(0)$ are thrown away. In this case, the additional, as compared to \eqref{Spln calc}, constant $\tilde{A}_2$ arising in Eq. \eqref{Spln calc new} must be zero.

As far as the real part of \eqref{I_II_gen} is concerned its dependence on a mass, and, in particular, vanishing of the term at $m^4\ln m$, is not influenced by a changing of a shape of the integration domain. However, the functions $g_i(y)$ change with $\De(x)$ analogously to the first term in \eqref{Spln calc 1}.

In the next section devoted to renormalization we shall analyze the one-loop contribution to the effective action in the form \eqref{Spln calc new} keeping in mind that a spoiling of conformal invariance modifies the massless contribution.

\section{Renormalization and analysis}\label{analysis}

In this section, we describe the renormalization procedure and give a brief analysis of possible physical effects following from the form of the obtained effective potential. A thorough study of the physical implications requires a detailed investigation of experimental data, modification of the obtained results to the standard model instead of the toy model \eqref{action} we used, and comparison with other approaches (see, e.g., \cite{DeWittDTGF,Hawk,FrolNov,GavGitdenmat}). It will be given elsewhere.

First of all, we take into account that the effective potential should reduce to the Coleman-Weinberg potential \cite{ColWein} on the spatial infinity $y=1$. This, in essence, results in a subtraction of values of the functions of $y$ in the effective Lagrangian \eqref{Spln calc} taken at $y=1$, and replacing all the coupling constants by their renormalized values. We shall not mark them by new labels and just bear in mind that such a renormalization has been done. Notice that the undefined constant $A_2$ is completely absorbed by this renormalization. The residuary constants $A_i$ have to be determined by some normalization conditions and, eventually, should be taken from experiments. As we have no any data on their values, we shall study possible scenarios for different $A_i$.

We start our investigation with the case $\tilde{A}_2=0$. Then, on substituting the masses of the particles in our model \eqref{W mass}, \eqref{Higgs mass} expressed in terms of $\eta$, the one-loop effective potential reads
\begin{multline}\label{eff pot real}
    V=\La+N^2A_0(1-y^{-2})+A_1\mu^2(1-y^{-1})+\frac{\mu^4}{64\pi^2}\ln(y|\mu^2+3\la\eta^2|)\\
    +\eta^2\left\{\frac{\mu^2}2\left[1+\frac{3\la}{16\pi^2}\ln(y|\mu^2+3\la\eta^2|)\right]+A_1[3(N-1)e^2+3\la](1-y^{-1})\right\}\\
    +\eta^4\left\{\frac\la4+\frac{3(N-1)e^4+9\la^2}{64\pi^2}\ln y+\frac{3(N-1)e^4\ln(e^2\eta^2)+9\la^2\ln|\mu^2+3\la\eta^2|}{64\pi^2}\right\},
\end{multline}
where $\La$ is the cosmological constant minus the value of the residual part of the potential taken in its minimum at the spatial infinity. For simplicity, we assume that the cosmological constant vanishes. Besides, the logarithms of the energy scale at which the coupling constants and masses are measured are included into $\La$, $\mu^2$, and $\la$. Because of that the expression for the effective potential \eqref{eff pot real} and some expressions below look wrong by dimensional reasons, but it is just an appearance.

In order to make our toy model being similar to the electroweak sector of the standard model, we choose $N=4$ and take the following values of the coupling constants and masses \cite{PartGroup2007} at $y=1$
\begin{equation}\label{coupl const}
    m_0=83,\qquad m_{H0}=150,\qquad\eta_0=247,\qquad\mu^2=-12445,\qquad e=\frac{m_0}{\eta_0},\qquad\la=\frac{m_{H0}^2}{2\eta_0^2},
\end{equation}
where the value of $\mu^2$ is taken so that the effective potential at the spatial infinity has a minimum at the experimental expectation value of the Higgs field $\eta_0$. Of course, choosing $N=4$ we have too many massless particles as compared to the electroweak sector of the standard model, but the difference in their number is irrelevant for our qualitative analysis.

On the Schwarzschild background, the one-loop effective action has an extremum when the field $\eta(y)$ satisfies an analog of the Ginzburg-Landau equation \cite{LandLifstat}
\begin{equation}\label{GL eqs}
    -y^{-1}(1-y)^4\left(\frac{d}{d\ln y}\right)^2\eta+V'(\eta)=0,
\end{equation}
where we already regard $\eta$ as the one component of the vector $\eta^n$, other components being zero. Besides, we neglect a self-interaction of the field $\eta$ through the gauge fields $A_\mu^a$. The kinetic term arising in \eqref{GL eqs} takes effectively into account the long-wave modes. Below we shall see that it is small on the functions minimizing the effective potential $V(\eta)$ in a sufficiently wide range of changing of the variable $y$ and can be considered as a perturbation. In that case, the below analysis is immediately applicable to an arbitrary static, spherically symmetric gravitational field.

In the conformal case, $\tilde{A}_2=0$, a crucial role for the physics near the black hole plays a sign of the constant $A_1$. Indeed, ignoring the kinetic term, we have
\begin{multline}\label{eta der}
    \eta'=-\frac{\partial_y V'}{V''(\eta)}=-\frac{2\eta}{y V''(\eta)}\\
    \times\left\{\frac{A_1}{y}[3(N-1)e^2+3\la]+\frac{3(N-1)e^4+9\la^2}{32\pi^2}\left[\eta^2+\frac{\mu^2}{3\la+(N-1)e^4\la^{-1}}\right]\right\}.
\end{multline}
The second term in the curly brackets is positive at the spatial infinity, because it vanishes at so small $\eta$ that the Higgs boson has an imaginary mass according to \eqref{Higgs mass}. Consequently, if $A_1\geq0$, the function $\eta(y)$ is monotonically decreasing since $V''>0$ at the minimum.

Assuming $\eta$ is finite at $y\rightarrow0$, it is not difficult to obtain from Eq. \eqref{GL eqs} without the kinetic term that either $\eta$ tends to zero or
\begin{equation}\label{eta asymp}
    \frac{A_1}{y}[3(N-1)e^2+3\la]\approx\frac{3(N-1)e^4+9\la^2}{32\pi^2}\left[\eta^2+\frac{\mu^2}{3\la+(N-1)e^4\la^{-1}}\right]\ln y.
\end{equation}
Therefore, $A_1$ should vanish, so does the term on the right hand side, but it is impossible as we have already seen above. Hence, the only two possibilities exist: $\eta$ tends to infinity or to zero at the horizon of the black hole.

At $A_1\geq0$, the function $\eta(y)$ is monotonically decreasing and tends to infinity at $y\rightarrow0$. To understand a behavior of $\eta$ at the negative $A_1$, we rewrite the minimum effective potential condition under the assumption $y\rightarrow0$ and $\eta\rightarrow\infty$ in the form
\begin{multline}
    -A_1[3(N-1)e^2+3\la]\\
    +y\eta^2\left[\frac\la2+\frac{3(N-1)e^4\ln e^2+9\la^2\ln(3\la)}{32\pi^2}+\frac{3(N-1)e^4+9\la^2}{32\pi^2}\left(\ln(y\eta^2)+\frac12\right)\right]\approx0.
\end{multline}
Hence, $\eta$ tends to infinity as $y^{-1/2}$. It is not hard to investigate the obtained asymptotic equation on $y\eta^2$. At $A_1<0$, it has two solutions, which coincide at $\bar{A}_1$, only if
\begin{multline}
    A_1\geq\bar{A}_1=-\frac{1}{32\pi^2}\frac{(N-1)e^4+3\la^2}{(N-1)e^2+\la}\\
    \times\exp\left[-\frac32-\frac{16\pi^2\la+3(N-1)e^4\ln e^2+9\la^2\ln(3\la)}{3(N-1)e^4+9\la^2}\right]\approx-4.66\cdot10^{-34}.
\end{multline}
The right dimension can be recovered if we recollect that $\la$ contains the logarithm of the energy scale at which the coupling constants are measured in experiments and take the values \eqref{coupl const}. The value of $\bar{A}_1$ depends on this scale through the coupling constants. Thus, at $A_1\in[\bar{A}_1,0]$, the Higgs field goes to infinity at the horizon, while at $A_1<\bar{A}_1$ it vanishes there.

Assuming $\eta$ to be small, we can expand $V'(\eta)$ around the point $\eta=0$ with the result
\begin{equation}\label{re Vpr}
     V'= A_1\left(1-\frac1{y}\right)[3(N-1)e^2+3\la]+\frac{\mu^2}{2}\left[1+\frac{3\la}{16\pi^2}\left(\ln|y\mu^2|+\frac12\right)\right]+\frac{3(N-1)e^4}{32\pi^2}\eta^2\ln\eta^2+\ldots
\end{equation}
We see that at $A_1<0$ and the sufficiently small $y$ there exists a value of the function $\eta$ that the written terms disappear. Equating the first two terms in \eqref{re Vpr} to zero, we can determine the critical point $y_0$, where $\eta$ vanishes. The derivative of the Higgs field $\eta'(y)$ at this point is infinity. In the narrow interval $A_1\in(\bar{A}_1,0)$, the real part of the effective potential has three extrema, two of them being minima. That is, in this interval the system may be found in a metastable state at small $y$. Notice that all the described features are typical for the Landau phase transition theory. As for the kinetic term, which we neglected in our considerations, it merely accelerates the process of going $\eta$ to zero or infinity.

So, we distinguish two regimes which can be called as the ``hot'' and ``cold''. In the  ``hot'' scenario, $A_1<0$, and the spontaneously broken symmetry is restored near the black hole. All particles falling on it become massless, the matter is disrupted and electrically charged particles radiate a huge number of high-energy photons which freely from the boundary of this region reach a remote observer. In the ``cold'' regime, $A_1\geq0$, the masses of all massive particles grow, when they approach the gravitating body. The high energy physics is suppressed by large masses, the matter rapidly gets cold and shrinks. The Bohr radius decreases, so does the range of action of the nuclear forces, while the energy level spacing increases.

Speculating in this way, we interpolate the one-loop results to a region where quantum corrections become large. Furthermore, the large gradients of $\eta$ produce a considerable gauge fields $A^a_\mu$ which act on the Higgs field and change it. Thus, we can believe in the obtained implications to a certain extent only, expecting that we catch a qualitative behavior of the Higgs field near a stationary, spherically symmetric gravitating body.

Now we turn to the case of a spoiled conformal invariance for massless modes $\tilde{A}_2\neq0$. It is not difficult to see even from \eqref{Spln calc new} that at $\tilde{A}_2>0$ the expectation value of the Higgs field grows exponentially near the horizon. Its asymptotics reads
\begin{equation}
    \ln(y\eta^2)\approx\frac{2\tilde{A}_2}{y}[3(N-1)e^4+9\la^2]-\frac{16\pi^2\la+3(N-1)e^4\ln e^2+9\la^2\ln(3\la)}{32\pi^2}.
\end{equation}
This is a highly unlikely scenario. At $\tilde{A}_2<0$, we have two regimes:
\begin{equation}
\begin{aligned}
    i)&\; \frac{A_1}{2\tilde{A}_2}&>A_1^s,\\
    ii)&\;\frac{A_1}{2\tilde{A}_2}&\leq A_1^s,
\end{aligned}
\end{equation}
where
\begin{equation}
    A_1^s:=-\frac{3\la\mu^2}{3(N-1)e^2+3\la}\approx4.39\cdot 10^3.
\end{equation}
In the first case, the phase transition occurs at the critical value $y_0$ determined by the equation
\begin{equation}\label{crit point}
    6\la\left(1-\frac1{y_0}\right)(A_1-2A_1^sA_2)=A_1^s\left[1+\frac{3\la}{16\pi^2}\left(\ln|y_0\mu^2|+\frac12\right)\right].
\end{equation}
The expectation value of the Higgs field has an infinite derivative in this point with respect to $y$. In the second case, the Higgs field tends to the finite nonzero value on the horizon
\begin{equation}
    \eta^2(0)=\frac{3(N-1)e^2+3\la}{3(N-1)e^4+9\la^2}\left(A_1^s-\frac{A_1}{2\tilde{A}_2}\right).
\end{equation}
For example, at vanishing $A_1/\tilde{A}_2$, quantum fluctuations of the Higgs field become tachyonic there. Thus, a breaking of the conformal invariance for massless modes may give rise to stabilization of the masses behaviour near the black hole.

\subsection{Normalization conditions}

As we have already mentioned, the constants $A_i$ should be determined from experiments. However, we can fix some of them demanding a fulfilment of certain properties from the Higgs field $\eta$ or the effective potential $V(\eta)$. As the effective potential $V$ is regular in $y$ at $y=1$, the field $\eta(y)$ is regular as well. Hence, $\eta$ approaches a constant value at the spatial infinity as $1/r$. It is possible to choose the constant $A_1$ so that $\eta$ tends to a constant faster, i.e., it goes to $\eta_0$ as $1/r^2$, keeping $\tilde{A}_2=0$. From Eq. \eqref{eta der} we deduce the corresponding value
\begin{equation}\label{Aw}
    A_1^{cr}=-\frac{3\la\mu^2+[3(N-1)e^4+9\la^2]\eta_0^2}{32\pi^2[3(N-1)e^2+3\la]}\approx-37.9.
\end{equation}
That is, we get to the ``hot'' case. If $A_1<0$ and $A_1>A_1^{cr}$ then the masses initially grow, when one approaches from the spatial infinity to the black hole, up to certain maximum values and then decrease down to zero. If $A_1<A_1^{cr}$, they monotonically tend to zero. For the marginal value \eqref{Aw} the Higgs field behaves at the spatial infinity like
\begin{equation}
    \eta\approx\eta_0-0.562(y-1)^2+\ldots
\end{equation}
In particular, assuming that the gravitational field in the outer space of the Sun changes slowly and is spherically symmetric, at least approximately, the relative mass shift on the Sun's surface, where $y-1\approx-4.24\cdot10^{-6}$, becomes
\begin{equation}
    \frac{\de\eta}{\eta_0}\approx-4.09\cdot10^{-14}.
\end{equation}
The critical point $y_0$ determined by Eq. \eqref{re Vpr} or \eqref{crit point} is found to be
\begin{equation}
    y_0\approx9.30\cdot10^{-3}.
\end{equation}
Recall that the kinetic term makes $y_0$ to be larger. For comparison we give the analogous quantities for the ``cold'' regime at $A_1=0$:
\begin{equation}
    \eta\approx\eta_0-1.12(y-1)+0.574(y-1)^2+\ldots\qquad\frac{\de\eta}{\eta_0}\approx1.92\cdot10^{-8}.
\end{equation}

Assuming $\tilde{A}_2\neq0$, we can make the masses of particles to be almost constant. For example, demanding the first two derivatives of $\eta(y)$ to be zero at the spatial infinity, we find
\begin{equation}
    A_1\approx2.89\cdot10^{10},\qquad\tilde{A}_2\approx-1.21\cdot10^6,\qquad \frac{\eta(0)-\eta_0}{\eta_0}\approx-4.80\cdot10^{-10}.
\end{equation}
However, the price we should pay for stability is the large energy density of the Higgs field, i.e., the large value of the effective potential taken in its minimum, which, moreover, is not integrable at the spatial infinity.

\begin{figure}[t]
\centering%
\includegraphics*[width=7cm]{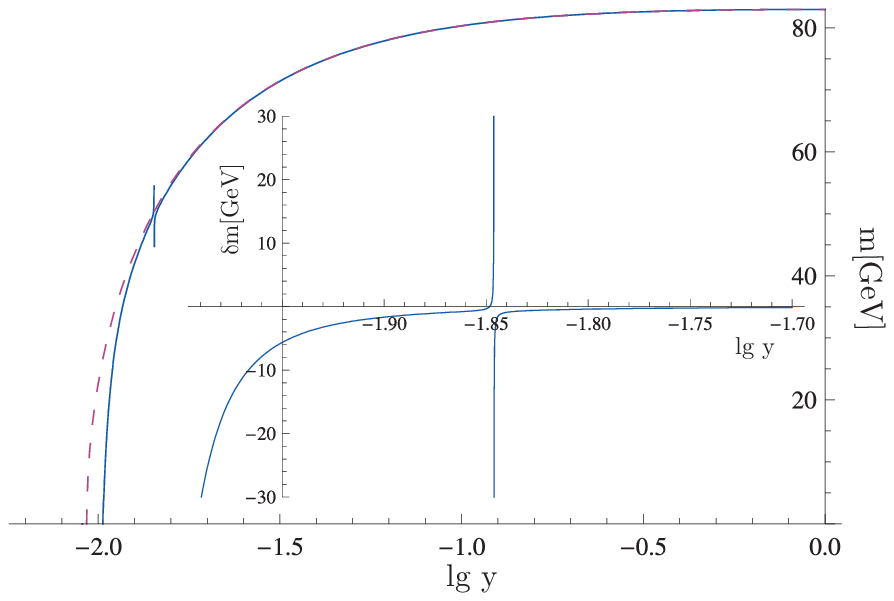}\;
\includegraphics*[width=7cm]{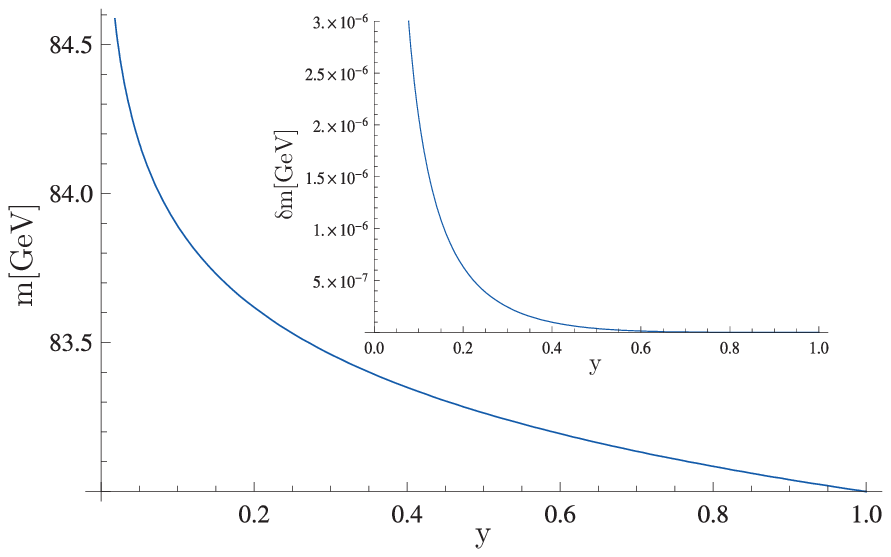}\\
\includegraphics*[width=7cm]{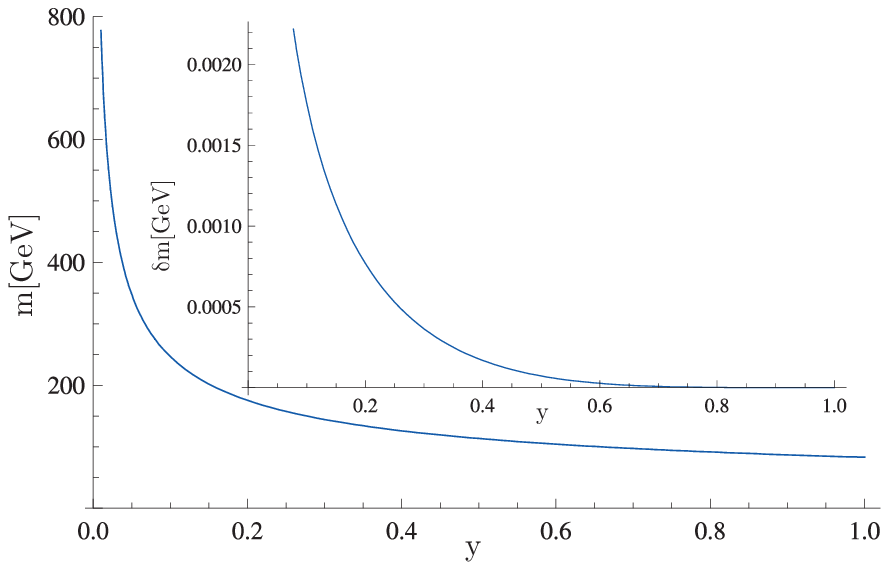}\;
\includegraphics*[width=7cm]{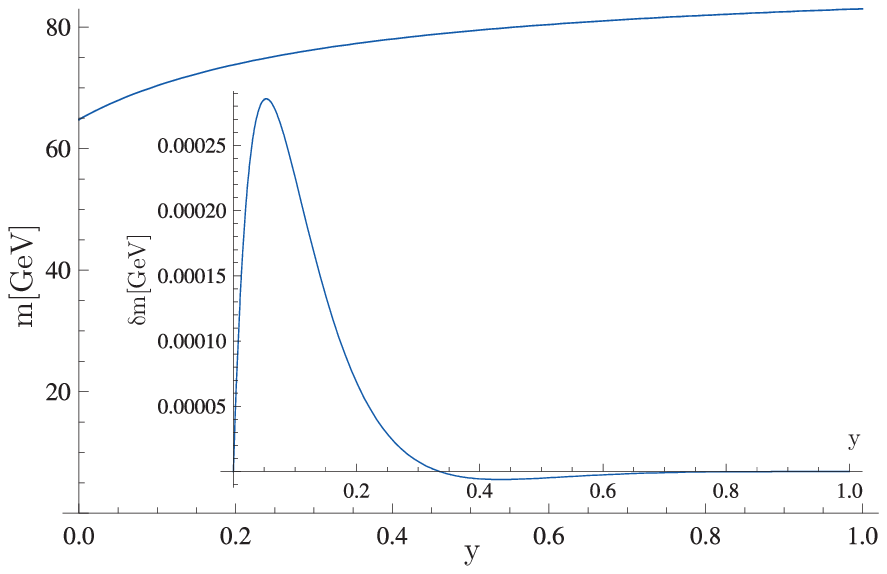}
\caption{{\footnotesize The dependence of a mass of the vector boson on the metric component $g_{00}\equiv y$. The small plots separately depict the corrections to the mass due to the kinetic term regarded as a perturbation. At the left on the top: the constant $A_1\approx-37.9$ as in Eq. \eqref{Aw} and $\tilde{A}_2=0$, the dashed line corresponding to the mass dependence with the kinetic term omitted. The discontinuity is at the point of a zero Higgs boson mass, where the effective potential is not smooth. At the right on the top: the constants $A_1$ and $\tilde{A}_2$ are taken to be zero. At the left on the bottom: the constants $A_1$ and $\tilde{A}_2$ are given in the first line of Eq. \eqref{Ai cold}. At the right on the bottom: the constants $A_1$ and $\tilde{A}_2$ are given in the second line of Eq. \eqref{Ai cold}.}}\label{plot of masses}
\end{figure}

Another reasonable normalization condition consists in the requirement that the energy density is integrable at the spatial infinity. To meet this requirement we should expand the potential in a Taylor series around the point $y=1$ up to $(y-1)^3$ order and assign its corresponding coefficients to be zero. At that, we have to exploit Eq. \eqref{GL eqs} without the kinetic term to express the derivatives of the Higgs field. After a little algebra, we arrive at the three equations on the three unknowns $A_0$, $A_1$, and $\tilde{A}_2$. They have two solutions
\begin{equation}\label{Ai cold}
\begin{aligned}
    i)&\;&A_0&\approx-9.36\cdot10^6,&\qquad A_1&\approx3.58\cdot10^3,&\qquad\tilde{A}_2&\approx1.04\cdot10^{-5},\\
    ii)&\;&A_0&\approx-1.01\cdot10^5,&\qquad A_1&\approx4.24\cdot10^2,&\qquad\tilde{A}_2&\approx-3.80\cdot10^{-2}.
\end{aligned}
\end{equation}
In the first case, masses of all massive particles grow exponentially in approaching the horizon. In the second case, the expectation value of the Higgs field decreases and tends to the finite nonzero value
\begin{equation}
    \frac{\eta(0)-\eta_0}{\eta_0}\approx-0.220.
\end{equation}
The asymptotics at the spatial infinity and the relative mass shifts on the Sun's surface are as follows
\begin{equation}
\begin{aligned}
    i)&\;&\eta&\approx\eta_0-107(y-1)+84.0(y-1)^2+\ldots&\quad V&\approx-1.4\cdot10^4 (y-1)^4+\ldots&\quad\frac{\de\eta}{\eta_0}&\approx1.84\cdot10^{-6},\\
    ii)&\;&\eta&\approx\eta_0+13.3(y-1)-10.4(y-1)^2+\ldots&\quad V&\approx3.65\cdot10^5 (y-1)^4+\ldots&\quad\frac{\de\eta}{\eta_0}&\approx-2.28\cdot10^{-7}.
\end{aligned}
\end{equation}
In spite of the requirement of integrability of the energy density at the spatial infinity entails that the conformal symmetry should be spoiled for massless modes, we
assume that the massless contribution to the one-loop effective potential is well approximated by the conformal invariant expression, which we derived. If the conformal symmetry for massless modes strictly holds, the energy integral of the Higgs field diverges at the spatial infinity, at least, logarithmically.

The plots of the masses $m(y)$, which are proportional to $\eta(y)$, and corrections to them due to the kinetic term in all these three cases are presented at Fig. \ref{plot of masses}.

%\newpage

\appendix

\section{Asymptotics of some integrals}

In this appendix, we consider asymptotics of the integrals, which we encounter calculating the effective potential. We need an asymptotic expansion for the four  integrals:
\begin{equation}
\begin{aligned}
    L_1&=\int_0^bdx\vf(x)\ln|x-\ups|,\\
    L_2&=\int_0^bdx\vf(x)\ln(x+\ups),\\
    M_1&=\int_0^bdx\vf(x)\frac{\ln|1-\frac{x}{\ups}|+\frac{x}{\ups}}{x^2},\\
    M_2&=\int_0^bdx\vf(x)\frac{\ln(1+\frac{x}{\ups})-\frac{x}{\ups}}{x^2},
\end{aligned}
\end{equation}
where $\ups$ tends to zero from the right and $\vf(x)$ is an integrable on $[0,b]$ function smooth at the point $x=0$. The first two integrals should be expanded up to $\ups^2$ and logarithmic corrections to it, while in the expansion of the integrals $M_1$ and $M_2$ we should retain the terms up to the finite part only.

The first integral reads
\begin{multline}
    L_1=\int_0^\ups dx\vf(x)\ln(\ups-x)+\int_\ups^bdx\vf(x)\ln(x-\ups)\\
    =\ups\int_0^1dx\vf(\ups x)[\ln\ups+\ln(1-x)]+\int_0^{b-\ups}dx\vf(x+\ups)\ln x.
\end{multline}
Expanding this in $\ups$ we arrive at
\begin{multline}\label{ap:j1 exp}
    L_1=\int_0^bdx\vf(x)\ln x-\ups\left[\vf(0)(1-\ln\ups)+\vf(0)\ln \e+\int_\e^bdx\frac{\vf(x)}{x}\right]\\
    -\frac{\ups^2}2\left[\vf'(0)\left(\frac12-\ln\ups\right)-\frac{\vf(0)}\e+\vf'(0)\ln \e+\int_\e^bdx\frac{\vf(x)}{x^2}\right]+o(\ups^2),
\end{multline}
where $\e$ tends to zero. Analogously, for the second integral we have
\begin{multline}\label{ap:j2 exp}
    L_2=\int_0^bdx\vf(x)\ln x+\ups\left[\vf(0)(1-\ln\ups)+\vf(0)\ln \e+\int_\e^bdx\frac{\vf(x)}{x}\right]\\
    -\frac{\ups^2}2\left[\vf'(0)\left(\frac12-\ln\ups\right)-\frac{\vf(0)}\e+\vf'(0)\ln \e+\int_\e^bdx\frac{\vf(x)}{x^2}\right]+o(\ups^2).
\end{multline}
The obtained parts of the asymptotic expansions of the integrals $L_1$ and $L_2$ pass into each other by the substitution $\ups\rightarrow-\ups$ and leaving $\ln\ups$ unchanged. Therefore,
\begin{multline}\label{ap:L}
    L:=\int_a^bdx\vf(x)\ln|x-\ups|=\int_a^bdx\vf(x)\ln|x|-\ups\left[\int_a^bdx\frac{\vf_1(x)}x+\vf(0)\ln\left|\frac{b}{a}\right|\right]\\
    -\frac{\ups^2}2\left[\int_a^bdx\frac{\vf_2(x)}{x^2}+\vf'(0)\ln\left|\frac{b}{a}\right|-\vf(0)\left(\frac1{b}-\frac1{a}\right)\right]+O(\ups^3),
\end{multline}
where $a<0$, $b>0$, and
\begin{equation}
   \vf_k(x):=\vf(x)-\sum_{n=0}^{k-1}\vf^{(n)}(0)\frac{x^n}{n!}.
\end{equation}

For the integral $M_1$ we have
\begin{multline}\label{ap:k1}
    M_1=\int_0^\ups dx\vf(x)\frac{\ln(1-\frac{x}{\ups})+\frac{x}{\ups }}{x^2}+\int_\ups^bdx\vf(x)\frac{\ln(\frac{x}{\ups}-1)+\frac{x}{\ups }}{x^2}\\
    =\frac1{\ups}\int_0^1dx\vf(\ups x)\frac{\ln(1-x)+x}{x^2}+\int_\ups^bdx[\vf_2(x)+\vf(0)+\vf'(0)x]\frac{\ln(\frac{x}{\ups}-1)+\frac{x}{\ups }}{x^2}.
\end{multline}
The second integral can be written to the required accuracy as
\begin{equation}
\begin{aligned}
    \int_0^{b-\ups}dx\frac{\vf_2(x+\ups)}{(x+\ups)^2}\ln\frac{x}{\ups}&=\int_0^bdx\frac{\vf_2(x)}{x^2}\ln\frac{x}{\ups}+o(1),\\
    \frac{\vf(0)}\ups\int_1^{b/\ups}dx\frac{\ln(x-1)+x}{x^2}&=-\frac{\vf(0)}\ups\left[\ln\frac{\ups}b+\frac\ups{b}\left(1-\ln\frac\ups{b}\right)\right]+o(1),\\
    \vf'(0)\int_1^{b/\ups}dx\frac{\ln(x-1)+x}{x}&=\vf'(0)\left(\frac{b}\ups+\frac12\ln^2\frac\ups{b}-1-\frac{\pi^2}6\right)+o(1).
\end{aligned}
\end{equation}
The first integral in \eqref{ap:k1} is easily expanded in $\ups$. Collecting all the terms together we obtain
\begin{multline}\label{ap:k1 exp}
    M_1=\frac1\ups\left[\int_0^bdx\frac{\vf_1(x)}x-\vf(0)\left(1+\ln\frac\ups{b}\right)\right]+\left[\frac{\vf(0)}b-\int_0^bdx\frac{\vf_2(x)}{x^2}\right]\ln\ups\\
    +\frac{\vf'(0)}2\ln^2\frac\ups{b}+\int_0^bdx\frac{\vf_2(x)}{x^2}\ln x-\frac{\pi^2}3\vf'(0)-\frac{\vf(0)}b(1+\ln b)+o(1).
\end{multline}
By the same way,
\begin{multline}\label{ap:k2 exp}
    M_2=-\frac1\ups\left[\int_0^bdx\frac{\vf_1(x)}x-\vf(0)\left(1+\ln\frac\ups{b}\right)\right]+\left[\frac{\vf(0)}b-\int_0^bdx\frac{\vf_2(x)}{x^2}\right]\ln\ups\\
    +\frac{\vf'(0)}2\ln^2\frac\ups{b}+\int_0^bdx\frac{\vf_2(x)}{x^2}\ln x+\frac{\pi^2}6\vf'(0)-\frac{\vf(0)}b(1+\ln b)+o(1).
\end{multline}
The expansions of the integrals $M_1$ and $M_2$ are very similar but not the same. Adding them, we can write
\begin{multline}\label{ap:M}
  M:=\int_a^bdx\vf(x)\frac{\ln|1-\frac{x}{\ups}|+\frac{x}{\ups}}{x^2}=\frac1{\ups}\left[\int_a^bdx\frac{\vf_1(x)}{x}+\vf(0)\ln\left|\frac{b}{a}\right|\right]\\
  -\ln\ups\left[\int_a^bdx\frac{\vf_2(x)}{x^2}+\vf'(0)\ln\left|\frac{b}{a}\right|-\vf(0)\left(\frac1{b}-\frac1{a}\right)\right]+\int_a^bdx\frac{\vf_2(x)}{x^2}\ln|x|\\
  +\frac{\vf'(0)}2\left(\ln\left|\frac{b}{a}\right|\ln|ba|-\pi^2\right)-\vf(0)\left(\frac1{b}-\frac1{a}+\frac{\ln b}b-\frac{\ln|a|}a\right)+o(1).
\end{multline}
Note that the terms proportional to $\ln^2\ups$ are canceled out.

\begin{acknowledgments}

I am indebted to A.A. Sharapov for reading of the draft of this paper, discussions and useful suggestions. The work is supported by the Russian Science and Innovations Federal Agency under contract No 02.740.11.0238, the RFBR grant 09-02-00723-a and the grant for Support of Russian Scientific Schools SS-871.2008.2.

\end{acknowledgments}

%\newpage


\begin{thebibliography}{999}


\bibitem{Hawk} S.~W. Hawking, \textsl{Particle creation by black holes}, Commun. Math. Phys. \textbf{43}, 199 (1975).

\bibitem{ColWein} S. Coleman, E. Weinberg, \textsl{Radiative corrections as the origin of spontaneous symmetry breaking}, Phys. Rev. D \textbf{7}, 1888 (1973).

\bibitem{Schwing} J. Schwinger, \textsl{On gauge invariance and vacuum polarization}, Phys. Rev. \textbf{82}, 664 (1951).

\bibitem{DeWittDTGF} B.~S. DeWitt, \textsl{The Global Approach to Quantum Field Theory. Vol. 1 and 2} (Claredon Press, Oxford, 2003).

\bibitem{VasilHeatKer} D.~V. Vassilevich, \textsl{Heat kernel expansion: user's manual}, Phys. Rep. \textbf{388}, 279 (2003), hep-th/0306138.





\bibitem{FaddSlav} L.~D. Faddeev, A.~A. Slavnov, \textsl{Gauge Fields: Introduction to Quantum Theory} (Benjamin-Cummings, New York, 1980).

\bibitem{HeTe} M. Henneaux, C. Teitelboim, \textsl{Quantization of Gauge Systems} (Princeton University Press, Princeton, New Jersey, 1992).

\bibitem{Wein} S. Weinberg, \textsl{The Quantum Theory of Fields. Vol. 2: Modern Applications} (Cambridge University Press, Cambridge, 1996).


\bibitem{BezShap} F. Bezrukov, M. Shaposhnikov, \textsl{Standard model Higgs boson mass from inflation: two loop analysis}, JHEP \textbf{07}, 089 (2009), arXiv: 0904.1537.

\bibitem{LyakhovichPhD} H. Leutwyler, \textsl{Gravitational field: Equivalence of Feynman quantization and canonical quantization}, Phys. Rev. \textbf{134}, B1155 (1964); E.~S. Fradkin, G.~A. Vilkovisky, \textsl{$S$-matrix for gravitational field. II. Local measure; general relations; elements of renormalization theory}, Phys. Rev. D \textbf{8}, 4241 (1973); S.~L. Lyakhovich, \textsl{The method of canonical quantization of theories with higher derivatives and its application in the $R^2$-gravity}, Ph.D. thesis, Tomsk State University, 1985 [in Russian].

\bibitem{BuchGit} I.~L. Buchbinder, D.~M. Gitman, \textsl{Calculating the probabilities of quantum processes in external gravitational fields. I}, Russ. Phys. J. \textbf{22}, 300 (1979); \textsl{Method of calculating probabilities of quantum processes in external gravitational fields. II}, Russ. Phys. J. \textbf{22}, 387 (1979).


\bibitem{MaslFed} V.~P. Maslov, M.~V. Fedoryuk, \textsl{Semi-Classical Approximation in Quantum Mechanics} (Reidel, Dordrecht, 1981).

\bibitem{BBT} V.~G. Bagrov, V.~V. Belov, and A.~Yu. Trifonov, \textsl{Methods of Mathematical Physics: Asymptotic Methods in Relativistic Quantum Mechanics} (TPU Publishing, Tomsk, 2006) [in Russian].

\bibitem{Valent} G. Valent, \textsl{Heun functions versus elliptic functions},  math-ph/0512006; P.~P. Fiziev, \textsl{Exact solutions of Regge-Wheeler equation and quasi-normal modes of compact objects}, Class. Quantum Grav. \textbf{23}, 2447 (2006), gr-qc/0509123; \textsl{Classes of exact colutions to the Teukolsky master equation}, arXiv: 0908.4234.

\bibitem{spectrum} I.~M. Ternov, V.~R. Khalilov, G.~A. Chizhov, and A.~B. Gaina, \textsl{Finite motion of massive particles in the Kerr and Schwarzschild fields}, Russ. Phys. J. \textbf{21}, 1200 (1978); I.~M. Ternov, A.~B. Gaina, and G.~A. Chizhov, \textsl{Finite motion of electrons in the field of microscopic black holes}, Russ. Phys. J. \textbf{23}, 695 (1980); D.~V. Gal'tsov, G.~V. Pomerantseva, and G.~A. Chizhov, \textsl{Behavior of massive vector particles in a Schwarzschild field}, Russ. Phys. J. \textbf{27}, 697 (1984).



\bibitem{FrolNov} V.~P. Frolov, I.~D. Novikov, \textsl{Black Hole Physics: Basic Concepts and New Developments} (Springer-Verlag, New York, 1998).


\bibitem{GSh} I.~M. Gel'fand, G.~E. Shilov, \textsl{Generalized Functions. Vol. I: Properties and Operations} (Academic Press, New York, London, 1964).

\bibitem{GavGitdenmat} S.~P. Gavrilov, D.~M. Gitman, and J.~L. Tomazelli, \textsl{Density matrix of a quantum field in a particle-creating background}, Nucl. Phys. B \textbf{795}, 645 (2008), hep-th/0612064.


\bibitem{PartGroup2007} W.-M. Yao \textit{et al}., \textsl{Review of particle physics}, J. Phys. G \textbf{33}, 1 (2006).

\bibitem{LandLifstat} E.~M. Lifshits, L.~P. Pitaevskii, \textsl{Statistical Physics. Part II} (Pergamon, New York, 1980).









\end{thebibliography}
\end{document}